\documentstyle[12pt]  {article}
\def\Journal#1#2#3#4{{#1}, {\bf #2} #3 (#4)}

\title{A symmetry test for quasilinear coupled
systems\footnote{appeared in Inverse Problems, 15 (1999), L5-L11}}
\author{V.V.Sokolov \\ Centre for Nonlinear Studies \& \\
Landau Institute for Theoretical Physics, \\ 
Kosygina 2, Moscow 117940, Russia \\ 
email: sokolov@landau.ac.ru  
\\ \\
Thomas Wolf\\ Queen Mary \& Westfield College, University of London, \\
Mile End Road, London E1 4NS, UK \\  email: T.Wolf@maths.qmw.ac.uk }

\def\phi{\varphi}
\def\be{\begin{equation}}
\def\ee{\end{equation}}
\def\ba{\begin{array}}
\def\ea{\end{array}}
\def\bea{\begin{eqnarray}}
\def\eea{\end{eqnarray}}
\def\bean{\begin{eqnarray*}}
\def\eean{\end{eqnarray*}}

\newfam\msbfam
\font\tenmsb=msbm10 
\textfont\msbfam=\tenmsb
\font\sevenmsb=msbm7 
\scriptfont\msbfam=\sevenmsb

\newfam\euffam
\font\teneuf=eufm10 \textfont\euffam=\teneuf
\font\seveneuf=eufm7 \scriptfont\euffam=\seveneuf

\begin{document}
\maketitle
\thispagestyle{empty}

\begin{abstract}
All quasilinear NLS-type systems with fourth order symmetry are
listed. Vector generalizations of some of them are constructed. Local
master symmetries for several systems from the list are found.
\end{abstract}


\section{Classification result.}

It is well known that the following class of systems of evolution equations
\bea
\label{nsgen}
\cases{
u_{t}=u_{xx}+F(u,v,u_x,v_x),\cr
v_{t}=-v_{xx}+G(u,v,u_x,v_x),\cr}
\eea
is very rich in integrable cases. 
In the papers \cite{MiSh1}-\cite{MiShYa2} by Mikhailov,
Shabat and Yamilov, all systems (\ref{nsgen}), possessing higher conservation
laws, were classified. Hence, these authors 
have found all systems (\ref{nsgen})
that can be integrated by the inverse scattering method
(S-integrable equations in the terminology by F. Calogero \cite{Cal}).

However, there are integrable cases that are not in
their classification. As an example, consider the system
\bea
\label{borov}
\cases{
u_t =   u_{xx} - 2 u u_x  - 2 v u_x  - 2 u v_x + 2 u^2  v +  2 u v^2,   \cr
v_t   =  - v_{xx} + 2 v u_x + 2 u v_x  + 2 v v_x - 2 u^2  v - 2 u v^2, \cr}
\eea
first discussed in \cite{BorPop} (see \cite{Sok} for generalizations). 
It can be reduced to
$$
U_t =   U_{xx}, \qquad
V_t   =  - V_{xx}
$$
by the following substitution of the Cole-Hopf type
\begin{equation}
u={U_x\over (U+V)},
\qquad v={V_x \over (U+V)}.
\label{borsub}
\end{equation}
The system (\ref{borov}) has no higher order conservation laws, but it has 
higher order symmetries. This is a typical feature of
linearizable systems like the Burgers equation
(C-integrable equations). Therefore,
it would be interesting to classify the systems (\ref{nsgen}), which have
higher order symmetries. As a result, all S-integrable and C-integrable
systems would be found.

The complete classification problem is very difficult. Here we consider
only the most interesting (from our opinion) subclass of systems (\ref{nsgen}).
Namely, we consider equations linear in all derivatives of the form
\bea
\label{kvazgen}
\cases{
u_t = u_{xx} + A_{1}(u,v) u_x + A_{2}(u,v) v_x + A_{0}(u,v) \cr
v_t = - v_{xx} + B_{1}(u,v) v_x + B_{2}(u,v) u_x + B_{0}(u,v). \cr}
\eea
without any restrictions on the functions $A_{i}(u,v), B_{i}(u,v).$ We apply 
to such systems the simplest version of the symmetry test 
(see \cite{IbSh1}-\cite{MiShSok}).

\paragraph{Lemma.} If system (\ref{kvazgen}) has a fourth order symmetry
\bea
\label{kvazsym}
\cases{
u_{\tau}=u_{xxxx}+f(u,v,u_x,v_x,u_{xx},v_{xx},u_{xxx},v_{xxx}),\cr
v_{\tau}=-v_{xxxx}+g(u,v,u_x,v_x,u_{xx},v_{xx},u_{xxx},v_{xxx})\cr}
\eea
then the system is of the following form
\bea
\nonumber
\cases{
u_t = u_{xx} + (a_{12} u v + a_1 u + a_2 v + a_0) u_x + (p_2 v +
p_{11} u^2 + p_1 u + p_0) v_x + A_{0}(u,v), \cr
v_t=- v_{xx} + (b_{12} u v + b_1 v + b_2 u + b_0) v_x + (q_2 u +
q_{11} v^2 + q_1 v + q_0) u_x + B_{0}(u,v), \cr}
\eea
where $A_0$ and $B_0$ are polynomials of at most fifth degree.

The coefficients of the last system satisfy an overdetermined
system of algebraic equations. The most essential equations are
\bea
\nonumber
\begin{array}{l}
p_2 (b_{12}-q_{11})=0, \qquad p_2 (a_{12}-p_{11})=0,
\qquad p_2 (a_{12}+2 b_{12})=0, \cr q_2 (b_{12}-q_{11})=0,
\qquad q_2 (a_{12}-p_{11})=0,
\qquad q_2 (b_{12}+2 a_{12})=0.
\end{array}
\eea
\bea
\nonumber
\begin{array}{l}
a_{12} (a_{12}-b_{12}+q_{11}-p_{11})=0, \qquad \ b_{12}
(a_{12}-b_{12}+q_{11}-p_{11})=0, \cr
(a_{12}-p_{11})(p_{11}-q_{11})=0, \qquad \qquad
(b_{12}-q_{11})(p_{11}-q_{11})=0, \cr
(a_{12}-p_{11})(a_{12}-b_{12})=0, \qquad
\qquad (b_{12}-q_{11}) (a_{12}-b_{12})=0.
\end{array}
\eea
As usual, such factorized equations lead to a tree of variants,
which was investigated by the computer algebra program 
{\sc Crack}\footnote{Wolf, T.\ and Brand, A.,
        The Computer Algebra Package CRACK for Investigating PDEs,
        manual + software in the REDUCE network library (1992)}
\cite{wolf2}.

Solving the overdetermined system, we don't consider so called
triangular systems like the following
\bea
\label{tri}
u_t=u_{xx}+2 u v_x, \qquad v_t=-v_{xx}-2 v v_x.
\eea
Here the second equation is separated
and the first is linear with the
variable coefficients defined by a given solution of the second equation.

\paragraph{Theorem.}
Any nonlinear nontriangular system (\ref{kvazgen}),
having a symmetry (\ref{kvazsym}),
up to scalings of $t$, $x$, $u$, $v$, shifts of $u$ and $v$, and the
involution
\be\label{inv}
u \leftrightarrow v, \qquad t \leftrightarrow -t
\ee
belongs to the following list:
\bea
\label{eq1}
\cases{
u_t=u_{xx} + (u + v) u_x + u v_x, \cr
v_t=-v_{xx}+ (u + v) v_x + v u_x, \cr}
\eea
\bea
\label{eq2}
\cases{
u_t= u_{xx} - 2 (u + v) u_x - 2 u v_x +2 u^2 v + 2 u v^2 +
\alpha u + \beta v + \gamma, \cr
v_t=-v_{xx} + 2 (u + v) v_x + 2 v u_x - 2 u^2 v - 2 u v^2  -
\alpha u - \beta v - \gamma, \cr}
\eea
\bea
\label{eq3}
\cases{
u_t=u_{xx} + v u_x + u v_x, \cr
v_t=-v_{xx} + v v_x + u_x, \cr}
\eea
\bea
\label{eq4}
\cases{
u_t=u_{xx}+2 v u_x+2 u v_x+2 u v^2 + u^2 +\alpha u + \beta v + \gamma, \cr
v_t=-v_{xx}- 2 v v_x - u_x, \cr}
\eea
\bea
\label{eq5}
\cases{
u_t=u_{xx}+\alpha v_x + (u + v)^2+ \beta (u+v) + \gamma, \cr
v_t=-v_{xx}+\alpha u_x - (u + v)^2 - \beta (u+v) - \gamma,    \cr}
\eea
\bea
\label{eq6}
\cases{
u_t=u_{xx} + (u + v) u_x + 4 \alpha v_x +
\alpha (u+v)^2 +\beta (u+v) +\gamma, \cr
v_t=-v_{xx} + (u + v) v_x + 4 \alpha u_x -
\alpha (u+v)^2 - \beta (u+v)-\gamma,  \cr}
\eea
\bea
\label{eq7}
\cases{
u_t=u_{xx}+2 \alpha u^2 v_x+2 \beta u v u_x  +
\alpha (\beta - 2 \alpha) u^3 v^2 + \gamma u^2 v+\delta u, \cr
v_t=-v_{xx}+2 \alpha v^2 u_x+2 \beta u v v_x  -
\alpha (\beta - 2 \alpha) u^2 v^3 - \gamma u v^2-\delta v, \cr}
\eea
\bea
\label{eq8}
\cases{
u_t=u_{xx} + 2 u v u_x + (\alpha + u^2) v_x, \cr
v_t=-v_{xx} + 2 u v v_x +(\beta + v^2) u_x,  \cr}
\eea
\bea
\label{eq9}
\cases{
u_t=u_{xx}+ 2 \alpha u v u_x + 2 \alpha u^2 v_x - \alpha \beta u^3 v^2
+\gamma u, \cr
v_t=-v_{xx}+2 \beta v^2 u_x + 2 \beta u v v_x + \alpha \beta u^2 v^3
-\gamma v, \cr}
\eea
\bea
\label{eq10}
\cases{
u_t=u_{xx}+ 2 u v u_x + 2 (\alpha + u^2) v_x+ u^3 v^2 + \beta u^3 +
 \alpha u v^2 +\gamma u, \cr
v_t=-v_{xx}-2 u v v_x - 2 (\beta + v^2) u_x- u^2 v^3-\beta u^2 v -
\alpha v^3  -\gamma v, \cr}
\eea
\bea
\label{eq11}
\cases{
u_t=u_{xx}+4 u v u_x+ 4 u^2 v_x + 3 v v_x + 2 u^3 v^2 + u v^3
+ \alpha u, \cr
v_t=-v_{xx} -2 v^2 u_x -2 u v v_x - 2 u^2 v^3 - v^4 - \alpha v, \cr}
\eea
\bea
\label{eq12}
\cases{
u_t=u_{xx}+4 u u_x + 2 v v_x, \cr
v_t=-v_{xx}-2 v u_x-2 u v_x- 3 u^2 v - v^3 + \alpha v, \cr}
\eea
\bea
\label{eq13}
\cases{
u_t=u_{xx}+v v_x, \cr
v_t=-v_{xx}+ u_x, \cr}
\eea
\bea
\label{eq14}
\cases{
u_t=u_{xx}+ 6 (u + v) v_x - 6 (u+v)^3-\alpha (u+v)^2-
\beta (u+v)-\gamma,
\cr v_t=-v_{xx}+ 6 (u + v) u_x + 6 (u+v)^3+\alpha (u+v)^2+
\beta (u+v)+\gamma, \cr}
\eea
\bea
\label{eq15}
\cases{
u_t=u_{xx}+v v_x, \cr
v_t=-v_{xx}+u u_x. \cr}
\eea

We omitted the term $(c u_x, c v_x)^T$ on the right hand sides of
all systems. It is a Lie symmetry, corresponding to the invariance
of our classification problem with respect to the shift of $x$.

\section{Discussion.}

{\bf 1. Admissible transformations.}
Some equations in the list contain arbitrary constants
$\alpha, \beta, \gamma, \delta$. Not all of them are essential.

Let us consider, for instance, the equations (\ref{eq8}). It is easy to see
that those constants $\alpha$ and
$\beta$ which are not equal to zero
can be reduced to 1 via scalings of $t,x,u$ and $v$.
In this way, (\ref{eq8}) actually describes three different
equations without parameters. These equations correspond to
$\alpha=\beta=1$, $\alpha=\beta=0$ and $\alpha=1, \beta=0$.

The following parameters: $\alpha$ in
(\ref{eq5}) and (\ref{eq6}), $\gamma$ in (\ref{eq7}), both
$\alpha$ and $\beta$ in (\ref{eq10}) are not essential in the same sense. For
(\ref{eq7}) and  (\ref{eq9}) the essential parameter is the ratio of
$\alpha$ and $\beta$. Note, that if $\alpha=\beta=0$ then
(\ref{eq7}) coincides with the nonlinear Schr\"{o}dinger equation, which is not
a separate equation in our list.

For some equations from the list, there exist admissible transformations of 
the form
\bea
\label{txtran}
u \rightarrow p(x,t)u + q(x,t), \quad v \rightarrow r(x,t)v +
s(x,t).
\eea
``Admissible'' means that the resulting equation does not depend
explicitly on $x$ and $t$ and has the same form (\ref{kvazgen}). 
Using such admissible transformations, one can remove some of constants 
in the equations of the list.

In particular, with the help of the transformation
$u \rightarrow \exp (c t) u,$ \
$v \rightarrow \exp (-c t) v$  one can remove the terms
$(c u, - c v)^T$ in (\ref{eq7}), (\ref{eq9}), (\ref{eq10}).

Using the transformations $u \rightarrow u + \lambda t +
\mu x,$ \  $v
\rightarrow v - \lambda t - \mu x$ it is possible to remove
$\beta$ and $\gamma$ in (\ref{eq5}), (\ref{eq6}). The equation
(\ref{eq14}) can be reduced to the form
\bea
\label{eq14'}
\cases{
u_t=u_{xx}+ 6 (u + v) v_x - 6 (u+v)^3 + c v_x,       \cr
v_t=-v_{xx}+ 6(u + v) u_x + 6 (u+v)^3 + c u_x       \cr}
\eea
by such a transformation and by shifts of $u$ and $v$. It seems to us that the
essential constant $c$ was missed in the classification result of \cite{ShYa}.

More general transformations are described in \cite{MiShYa1},\cite{MiShYa2}
which reduce some of the equations (\ref{eq1})-(\ref{eq15})
to others in this list. For simplicity in applying our results,
we will not rely on these non-trivial
transformations, and will instead operate with the complete
list (\ref{eq1})-(\ref{eq15}).  

{\bf 2. Three groups of equations.} All equations of the list can 
be divided into three groups. 
The first group contains the so called NLS-type equations (\ref{eq1}),
(\ref{eq3}),(\ref{eq6}),(\ref{eq7}),(\ref{eq8}). Besides a higher symmetry 
(\ref{kvazsym}) every such equation possesses a symmetry of the form
\bea
\label{sym3}
\cases{
u_{\tau}=u_{xxx}+\phi (u,v,u_x,v_x,u_{xx},v_{xx}),\cr
v_{\tau}=v_{xxx}+\psi (u,v,u_x,v_x,u_{xx},v_{xx}).\cr}
\eea
This is typical for equations having the Lax representations in $sl(2)$.

The equations of the Boussinesq type form the second group 
(\ref{eq5}),(\ref{eq13}),(\ref{eq14}),(\ref{eq15}). 
They have no  symmetries of
third order. This indicates the existence of a Lax representation in
$sl(3)$. We have chosen the existence of the symmetry
(\ref{kvazsym}) as a criterion of integrability for (\ref{kvazgen})
since the choice of the simplest ansatz (\ref{sym3}) leads to the loss of all 
equations of the second group.

The last group
(\ref{eq2}),(\ref{eq4}),(\ref{eq9}),(\ref{eq10}),(\ref{eq11}),
(\ref{eq12}) consists of "linearizable" equations, which have no
higher conservation laws. Some of them seem to be new.

{\it Equations (\ref{eq9}), (\ref{eq10}) :}
In \cite{EkCal},\cite{Cal} one can find a linearization
procedure for (\ref{eq9}). 
Namely, a non-local substitution 
\be\label{ekha}
U=u \exp{\left(\alpha \int uv\, dx\right)}, 
\qquad V=v \exp{\left(-\beta \int uv\, dx\right)}
\ee
reduces it to the linear equation $U_t=U_{xx}+\gamma U, \ 
V_t=-V_{xx}-\gamma V.$

Consider equation (\ref{eq10}).
It is easy to see that it has the following symmetry 
\bea\label{ibsh2}
\cases{
u_{\tau}=u_{xxx}+3 u v u_{xx}+6 u u_x v_x+3 v u_x^2+3 u^2 v^2 u_x,\cr
v_{\tau}=v_{xxx}+3 u v v_{xx}+6 v u_x v_x+3 u v_x^2+3 u^2 v^2 v_x,\cr}
\eea
for any $\alpha, \beta, \gamma.$ Under a reduction $u=v$, (\ref{ibsh2}) 
coincides with the well known equation (see \cite{IbSh2},\cite{SokSh},
\cite{Cal})
$$
u_{\tau}=u_{xxx}+3 u^2 u_{xx}+9 u u_x^2+3 u^4 u_x,
$$
which can be linearized by the substitution $U=u \exp{\int u^2 dx}$. 
It is easy to verify that the following very similar substitution 
(cf. also with (\ref{ekha}))
\be\label{subibsh2}
U=u \exp{\left(\int u v\, dx\right)}, 
\qquad V=v \exp{\left(\int u v\, dx\right)}
\ee
reduces (\ref{ibsh2}) to $U_{\tau}=U_{xxx}, \ V_{\tau}=V_{xxx}$. 
The same substitution reduces (\ref{eq10}) to a linear system 
$$
U_t=U_{xx}+2 \alpha V_x+\gamma U, \qquad V_t=-V_{xx}-2 \beta U_x-\gamma V.
$$ 

Generalizing the formula (\ref{subibsh2}) one can find the following 
vector generalizations of the systems (\ref{eq10}) and (\ref{ibsh2}):
\bea
\nonumber
\cases{
u_{t}=u_{xx}+ 2 <u, v> u_{x} + 2 <u, v_x> u + <u,v>^2 u + \cr 
\qquad 2 \alpha v_x+\beta <u, u> u+ 2 \alpha <u, v> v - \alpha <v, v> u+
\gamma u,
\cr
\quad \cr
v_{t}=- v_{xx} - 2 <u, v> v_{x} - 2 <v, u_x> v - <u,v>^2 v - \cr
\qquad 2 \beta u_x-\alpha <v, v> v - 2 \beta <u, v> u + \beta <u, u> v-
\gamma v,
}
\eea

\bea
\nonumber
\cases{
u_{\tau}=u_{xxx}+3 <u, v> u_{xx}+3 u <u_x, v_x>+3 <u,v>_x u_x+3 <u,v>^2 u_x,\cr
v_{\tau}=v_{xxx}+3 <u, v> v_{xx}+3 v <u_x, v_x>+3 <u,v>_x v_x+3 <u,v>^2 v_x,\cr}
\eea

where $u$ and $v$ are $N$-dimensional vectors and $< \ >$ is a scalar product. 
Both of them can be linearized just as in the scalar case: 
\be
U=u \exp{\int <u, v> dx}, \qquad V=v \exp{\int <u, v> dx}.
\ee

In contrast with (\ref{eq9}), (\ref{eq10}), equations (\ref{eq2}) and 
(\ref{eq4}) are related to linear systems 
\be\label{lin}
U_t=U_{xx}+c_1 U_x+c_2 V_x+c_3 U, \qquad 
V_t=-V_{xx}-k_1 V_x-k_2 U_x-k_3 V
\ee
via {\bf local} differential substitutions. 

{\it Equation (\ref{eq2}):}  For 
$\alpha=\beta=\gamma=0$ the substitution is given by (\ref{borsub}). In the 
general case such a substitution is defined by 
$$
u={U_x\over (U+V)}+{(c_1-k_2) U+c_2 V\over 2 (U+V)},
\qquad v={V_x \over (U+V)}+{(c_1-c_2) V+k_2 U\over 2 (U+V)},
$$
where the constants in (\ref{lin}) and (\ref{eq2}) satisfy the following 
conditions
$$
k_1=c_1, \qquad k_3=c_3={c_1 (c_1-c_2-k_2)\over 4}, 
$$
$$
2 \alpha =-k_2 (c_1+c_2), \qquad 2 \beta =-c_2 (c_1+k_2), \qquad 
2 \gamma =c_1 k_2 c_2. 
$$

{\it Equation (\ref{eq4}):}
The substitution is of the form 
$$
u={c_2 U_x\over V}+{c_2^2\over 2},
\qquad v={V_x / V}+{c_1\over 2}.
$$
The relations between constants are the following
$$
k_1=c_1, \qquad k_2=c_2, \qquad k_3=0, 
$$
$$
2 \alpha =-c_1^2-2 c_2^2+2 c_3, \qquad \beta =-c_1 c_2^2, \qquad 
4 \gamma =c_2^2 (2 c_1^2 + c_2^2-2 c_3). 
$$

{\it Equation (\ref{eq11}), (\ref{eq12}):} 
These equations are related to triangular systems.  
Equations (\ref{eq12}) have been obtained in \cite{OlSok}. 
The substitution
$$
u = {U_x \over 2 U}, \qquad v = {V \over \sqrt {U}};
$$
found first by Marihin \cite{Mar},
links equations (\ref{eq12}) with
$$
U_t= U_{xx} + 2 V^2,  \qquad V_t= - V_{xx} + \alpha V.
$$
The last system is linear in the 
following sense. To find $V$ we need to solve a linear equation. 
For a given function $V$, the function $U$ satisfies a linear 
equation with variable coefficients.

Similarly the following substitution
$$
u={1 \over 3} U^{-2/3} V^{-1} U_x, \qquad v=U^{-1/3} V,
$$
reduces (\ref{eq11}) to
$$
U_t=U_{xx}-2 V^{-1} V_x U_x+3 \alpha U+3 V^3, \qquad V_t=-V_{xx}.
$$

{\bf 3. Master-symmetries.} For all equations 
(\ref{eq1})-(\ref{eq15}) we have found all
symmetries of order less or equal than four using 
the computer program {\sc LiePde} \cite{LiePde}. It turns
out that many of the equations have symmetries depending on $t$ and $x$
explicitly. 
For example, computing for equation (\ref{eq9}) the general symmetry of 
$n^{\mbox{th}}$ order of the form 
$$
u_{\tau}=P(t) u_{nx} + \ldots,  \qquad
v_{\tau}=-P(t) v_{nx} + \ldots
$$
the polynomial $P(t)$ is an arbitrary polynomial of degree $n$.

It is well known that symmetries which are linearly depending on $t$ and
$x$ are closely related to local master-symmetries \cite{Fu}. The
equations (\ref{eq1})-(\ref{eq4}),
(\ref{eq8})-(\ref{eq10}) have such symmetries. To obtain the
master-symmetries one has to simply put $t$ equal to zero in these 
time-dependent symmetries. 
 
The resulting master-symmetry is of the form
\bea
\label{mast}
\cases{
u_{\tau}=2 x \ (u_{xx}+F(u,v,u_x,v_x))+f(u,v,u_x,v_x),\cr
v_{\tau}=2 x \ (-v_{xx}+G(u,v,u_x,v_x))+g(u,v,u_x,v_x),\cr}
\eea
where $F$ and $G$ are the right hand sides of the corresponding
equation (\ref{nsgen}) and $f$ and $g$ are given by the following
list
\bea
\nonumber
\begin{array}{rll}
(\ref{eq1}): & f=u^2+3 u v+4 u_x, & g=v^2+3 u v - 4 v_x, \cr
(\ref{eq2}): & f=2 u^2+4 u v+\beta+ 3 u_x, & g=-2 v^2-4 u v-\alpha - 3 v_x,\cr
(\ref{eq3}): & f=4 u v+5 u_x, & g=v^2+4 u - 5 v_x, \cr
(\ref{eq4}): & f=4 u v+\beta+3 u_x, & g=-2 v^2-2 u - \alpha - 3 v_x,\cr
(\ref{eq8}): & f=2 u^2 v+2 \alpha v+3 u_x, & g=2 u v^2+2 \beta u - 3 v_x,\cr
(\ref{eq9}): & f=2 \alpha u^2 v-2 \gamma x u+2 u_x, & g=2
                                     \beta v^2 u +2 \gamma x v - 2 v_x,\cr
(\ref{eq10}): & f=2 u^2 v+2 \alpha v+2 u_x, & g=-2 u v^2-2 \beta u - 2 v_x.
\end{array}
\eea

\section*{Acknowledgement}
The first author (V.S.) was supported, in part, by RFBR grant 99-01-00294, 
EPSRC grant GR/L99036 and LMS(fSU) grant 5324. He is grateful to the 
Queen Mary \& Westfield College (Univ.\ of London) for its hospitality.
The second author (T.W.) thanks the Symbolic Computation Group at the
University of Waterloo for its hospitality during the authors
sabbatical in the fall of 1998.


\end{document}